\documentclass[runningheads]{llncs}
\usepackage{listings}
\lstdefinelanguage{json}{
    basicstyle=\ttfamily\footnotesize,
    showstringspaces=false,
    breaklines=true,
    breakatwhitespace=false,
    columns=fullflexible,
    keepspaces=true,
    frame=single,
    tabsize=2,
}

\lstdefinestyle{prompt}{
    basicstyle=\ttfamily\small,
    breaklines=true,
    breakatwhitespace=false,
    columns=fullflexible,
    keepspaces=true,
    showstringspaces=false,
    frame=none
}

\usepackage[T1]{fontenc}
\usepackage{graphicx}
\usepackage{multirow}
\usepackage{pdflscape}
\usepackage{placeins}
\usepackage{tikz}
\usepackage[most]{tcolorbox}
\usepackage{subcaption}
\usepackage{amsmath}
\usepackage{hyperref}
\usepackage{url}
\usepackage{booktabs}
\usepackage{array}
\usepackage{ragged2e}
\usepackage{threeparttable}
\usepackage{makecell}

\usepackage{xcolor}
\usepackage{colortbl}

\definecolor{tableheadbg}{RGB}{30, 60, 114}     
\definecolor{tablerowalt}{RGB}{240, 245, 252}   
\definecolor{tablerowwhite}{RGB}{255, 255, 255} 

\usepackage[authoryear,longnamesfirst]{natbib}

\usepackage{pgfplots}
\pgfplotsset{compat=1.18}
\usepackage{pgfplotstable}

\usepgfplotslibrary{statistics}
\begin{document}
\title{ReCon: A Resource-Constrained Benchmark for LLM-Based Cybersecurity Compliance Across Ingestion and Retrieval Pipelines}
\titlerunning{ReCon}
%
\author{Rohit Negi\inst{1}\orcidID{0000-0002-4211-5637} \and
Rishik Jain\inst{2}\orcidID{0009-0008-6267-180X} \and
Soumyo V Chakarborty\inst{2}\orcidID{0009-0008-8078-0196} \and
Amit Negi\inst{2}\orcidID{0009-0005-6016-5725}\and
Sandeep K Shukla\inst{2}\orcidID{0000-0001-5525-7426}}
\authorrunning{Negi et al.}
%
\institute{Indian Institute of Technology Kanpur, India \and
International Institute of Information Technology Hyderabad, India
\email{rohit@cse.iitk.ac.in}
}
\maketitle              
\begin{abstract}
With the increasingly aggressive cyber threat landscape for governments, businesses, and institutions, as information and/or cyber security implementations are increasingly under scrutiny by regulators, it has been pointed out that governance failure is one of the major reasons for a weakened cyber security posture. A major component of Cyber/information security governance is the development, adoption, and implementation of a comprehensive information and/or cyber security policy document. The policy document must be in compliance with international or national standards and, if possible, with regulatory guidelines.    However, it is often observed that policy documents are often incomplete with respect to industry standards or regulations and  require revision when subjected to a thorough audit. Identifying the gaps between the controls and processes documented in the policy and those required in the regulations or standards  necessitates  extensive manual effort.    The advent of Generative AI tools such as Large Language Models (LLMs) led to use of LLMs and Agentic AI tools to automate such compliance checks, as seen in a few research publications in recent times. However, such reported use of LLMs are experimented with high resource environments such as expensive  GPUs and memory based servers.  For smaller organizations such expensive compute platform may not be easily available. In addition, policy documents are often changed and improved and every iteration would require a compliance assessment.  Therefore, it would be helpful if such compliance checks can be done as often as required without paying high price of GPU and memory expensive compute. In this article, we benchmark the compliance checking tasks on LLMs that do not require GPU and high memory usage and the effectiveness of such resource constrained LLMs in compliance checking. Our experiments demonstrated that the low resource LLMs can provide good agreement/accuracy  in compliance checking of policy documents against standards by experimenting with ISO 27002:2022 controls against multiple policy documents.
\end{abstract}
\section{Introduction}

Increasing use of digitalization of power grid, manufacturing, transportation systems, communication systems, banking and financial organizations, government institutions has been extending the cyber attack surface, rapidly as evidenced by a large number of cyber attacks every year in all sectors of the economy. The extensive use of 5G communication, generative and agentic AI based tools and Industry 4.0 has further aggravated   the big cybersecurity challenge \cite{chaudhuri2025factors}. In various studies, weak information and/or cyber security governance and a lack of skilled workers have been found to contribute greatly to the weak cyber security posture. For example, the FICCI-EY 2026 risk survey shows that Indian companies are increasingly concerned about cybersecurity\cite{ey2026risk}. This survey further reveals that effective governance pressures are intensifying, with 45\% of respondents indicating that non-compliance with governance  requirements has had  a direct impact on their organizations. In addition, workforce challenges persist, as 64\% of organizations report that there are talent shortages and skill gaps that affect performance and long-term capability planning. At the governance level, organizations implement structured frameworks such as the NIST cybersecurity framework\cite{NIST2024CSF} and/or the ISO/IEC 27001 standard\cite{ISO27001} to standardize risk identification, assessment and control processes. These frameworks help organizations establish information/cyber security policies in accordance with these standards. Sector specific regulators often require that regulated entities document and implement policies based on sector specific guidelines or widely adopted standards such as NIST SP 800-53 or ISO 27002 etc. The policy documents also define roles and responsibilities and ensure continuous monitoring and compliance. Many countries also provide guidance in the form of policies\cite{certin_advisory_170}, and guidelines \cite{irdai_information_cyber_security_guidelines_2023} \&  \cite{certin_guidelines_list} \cite{certin_guidelines_govtentities_2023}, which are  issued from time to time  to reinforce cybersecurity management. However, a recent  survey \cite{HAUFE2016339} with 90 experts was conducted to identify the ISMS core processes and unfortunately ``the information security governance process" was not clearly identified as a core process. In \cite{Oh2025} consolidated standards and best practices and identified that organizations can use structured processes from the Strategic Enterprise Cybersecurity Governance Model (ECyG-M) to identify, evaluate, prioritize and manage cyber risks to achieve their strategic goals. NIST realized that cybersecurity is not just a technical exercise, it’s fundamentally an organizational responsibility. That’s why in CSF 2.0 \cite{NIST2024CSF}, there is an explicit function for governance, leadership, risk oversight, policy enforcement; the overarching influence across all functions. Cyber risk also arise internally, due to employee errors, insider threats, or lapses in security policies and processes, and not just due to technical inadequacy. In \cite{JAEGER2021103318}, it has been mentioned that a key instrument for reducing the threats to information security associated with employees is to create, deploy, and enforce information security policies. In addition to this, \cite{VANCE2012190} also argues that employees who do not comply with IS security procedures are a key concern for organizations today. In summary, it is amply clear that  cyber security works  effectively only if governance is strong and strong  technology based controls alone do not suffice. 

To mitigate cyber risks, organizations must implement a mature \cite{doe_c2m2} Information/Cyber security Management System (ISMS/CSMS) that covers assets, configuration, vulnerability, patch and control management, as well as audits, incident response and recovery. Equally important are the effectiveness and quality of the ISMS/CSMS. The ISO standards \cite{ISO27001} identify nine key elements for evaluation: planning, leadership, risk management, policy management, resource management, communication, management review, documentation, and audit. CERT-In, a leading regulatory body in India, has mentioned in its guidelines \cite{certin_guidelines_govtentities_2023} for government organizations that internal cyber audit should be conducted twice a year and third party audit should be conducted at least once a year. However, A cyber audit is a time-consuming and participatory process - it requires the active participation and time of not only the cyber security department but also other departments within the organization. All audit activities must be aligned with organizational policies that provide the framework and guidance for consistent and reliable information/cyber security operations. In essence, well-documented policies are critical to ensuring a resilient and high-performing ISMS/CSMS.

The methodology used to audit an organization's ISMS/CSMS  may vary from one organization to another. Some may adopt standard frameworks such as ISO 27001 \cite{ISO27001} or NIST CSF \cite{NIST2024CSF}, while others may use customized approaches tailored to their specific needs \cite{11416280}. According to industry best practices, an effective ISMS/CSMS follows three repeating phases: Define, Implement, and Manage. First, the organization defines its information/cyber security approach through policies approved and communicated by the top management. Next, these policies are implemented through processes, technical and administrative controls, along with training and awareness programs. However, some aspects may be missed during the initial definition, and the requirements may change over time due to new regulations, standards, or evolving threats. In the management phase, ISMS/CSMS is reviewed against updated requirements to identify gaps. These gaps are then addressed in the next cycle by refining policies and controls. This continuous and iterative process ensures that cybersecurity is up to date with changing risks, technologies, and regulatory expectations. In a ISMS/CSMS, auditors, auditees, and regulatory bodies all play important roles in ensuring compliance. The review process helps organizations identify gaps and improvement opportunities, supports auditors in assessing business continuity, and enables regulators to evaluate how well standards and guidelines are being followed across organizations and sectors. This requires all three stakeholders to carefully analyze the policy documents in accordance with relevant frameworks and regulations.

However, this process is often challenging due to limited availability of trained human resources\cite{ey2026risk} and the time required to review extensive documentation. For example, ISO 27002:2022 contains a total of 93 controls. Policy documents are typically written in a subjective and complex manner, making them subject to human interpretation, especially when the human auditor is inexperienced or has language difficulties. As a result, auditors, auditees, and regulators face a high cognitive burden when trying to extract meaning, context, and insight from large volumes of text during assessments. In \cite{chaudhuri2025factors}, researchers show that cybersecurity technology, staff confidence in cybersecurity, and processes all influence cybersecurity transformation. Therefore, businesses need a clear information and cybersecurity policy and the right tools to support review and audit processes. Similarly, \cite{ey2026risk} shows that 59\% of respondents to their survey believe that the limited adoption of emerging technologies such as AI hinders operational effectiveness.


Given the scarcity of experienced human resources, especially in  the small- and medium scale industry and organizations, the creation, update, review, and compliance check against regulatory requirements is resource expensive and consequently ignored by such organizations. It is therefore desirable that a resource economic and  automated compliance check tool be made available that could either be applied against a policy document to generate a compliance report or as assistant to the human expert. This will reduce the time spent reviewing policy documents. In this way, the current capacity gap could be addressed significantly. 

In the recent past, quite a few tools have been discussed in the literature which use agentic AI for automating audit and compliance tasks. Most of these agents run on LLMs, and report that API calls to closed source LLMs do quite well.  However, not all organizations can afford the cost of too many API calls or token costs. Most medium- and small scale organizations do not possess their own high-capacity infrastructures, such as GPUs and servers with large memory. The target users for this work are the medium- and small scale organizations who need low cost cyber security to protect their data, better cyber hygiene of their employees, administrative controls for securing business processes and data flow, and possibly open source models and tools for securing their infrastructure. This work investigates whether policy conformity and violations analysis leveraging open source LLMs can be carried out effectively without having to procure expensive GPU servers or high-memory configuration systems.  This work analyzes multiple open weight LLMs and various configurations for policy compliance task and determines the configuration and LLMs that can be used such that the resource usage is within the budget of small/medium organizations. For the purpose of this analysis, conformity refers only to documentary coverage of the control requirements and not to operational implementation or effectiveness.

In this paper, we make the following main contributions:

\begin{enumerate}

    \item We demonstrate that even under resource constraints such as unavailability of GPU, and low memory, one could run open-weight LLMs with appropriate RAG architecture to get reasonably acceptable agreement in automated compliance checking for organizational information/cyber security policies against standards or guidelines. 
    
    \item We carry out a series of experiments with 8 well known open weight LLM models for the compliance check automation. We evaluate the agreement and disagreement  against human expert determined check for compliance. Such experiments help us choose appropriate LLM for the task. 

    \item We carry out a series of experiments with 2 different vector embedding models, 2 different re-ranking models and different retrieval configurations for the best agreement score. Such experiments help us choose appropriate embedding/re-ranking model, and retrieval configurations for the task. 
    
    \item We also  benchmark different LLM and configurations  in terms of latency, throughput, CPU utilization and memory usages while assessing agreement and disagreement. 
    
\end{enumerate}

\subsection{Organization}
This paper is organized as follows. Section 2 presents a comprehensive review of the existing literature, focusing on prior work related to LLMs, and their applications in cybersecurity governance. Section 3 describes the methodology adopted in this study, including the overall system architecture, pre-processing such as chunking, intelligent retrieval, and LLM-based reasoning. Section 4 describes the experiment, detailing the computational environment, datasets used, retrieval strategies for prompt construction, and combinations of runs. Section 5 presents the results obtained from the experiment. Finally, Section 6 concludes the paper by summarizing the key findings, limitations and directions for future research.

\section{Literature Review}

Cybersecurity compliance and audits have traditionally been labor - intensive processes that often involve a degree of subjectivity. Auditors are required to interpret relevant standards, examine organizational policies, and assess whether security controls have been properly implemented. During a compliance assessment, regulatory bodies ensure that the organization has properly adopted and implemented the guidelines and standards they set. Furthermore, auditee conducts an internal audit, it also has to review various policies, procedures, records and other related documents. LLMs are increasingly being investigated for applications in governance, risk, and compliance (GRC) and have significantly transformed ISMS/CSMS practices  \cite{11129535} \cite{tang2023policygpt}  \cite{WOODRING2024103997} \cite{saha2025parag} \cite{11204572} \cite{lodge2024rage} \cite{MCINTOSH2024103964}. As policy review is time-consuming, One can either use commercial tools such as ServiceNow GRC \cite{servicenow}, MetricStream \cite{metricstream}, etc., or opt for policy compliance using closed source LLM's\cite{negi2026automating}. AI-driven solutions rely on API-based services with token-based pricing models, which can become costly when deployed at scale. 

Agentic AI-based solutions \cite{negi2026automating} have demonstrated satisfactory accuracy across various tasks. However, many organizations are hesitant to share their policies and sensitive documents with cloud-based closed-source models. Therefore, they often prefer on-premises solutions, which allow them to maintain complete control over the data. Running such solutions on-premises using dedicated GPUs entails substantial upfront hardware investments as well as ongoing operational and maintenance expenses.

A conceptual framework has demonstrated that BART-based semantic analysis can be used for policy compliance assessment \cite{priescu2026automated}. However, this approach assumes that the auditee has access to the required ISO standard documents. In practice, this may not always be the case, as access to these documents often requires a paid subscription. This raises an important question: Is it possible to conduct policy compliance analysis using LLMs prior knowledge when the auditee does not have access to the full ISO standards?

If a model contains sufficient prior knowledge of auditing or compliance standards, repeatedly incorporating detailed guidelines into the prompt can result in unnecessary token expenditure. In this context, In \cite{negi2026small} present an evaluation and comparative analysis of various low-computation open-weight models. The study tested 16 open-weight models and one closed-weight model, demonstrating that many open-weight models also possess sufficient knowledge of compliance and auditing topics.  Table \ref{tab:llm_compliance_comparison}  summarizes existing research on the use of LLMs and AI in the field of policy. These studies aim to understand how AI-based techniques can make policy formulation, document analysis, compliance assessment, and auditing processes more effective, accurate, and automated.

\begin{table}[]
\centering
\caption{Comparative Analysis of LLM-based Approaches for Cybersecurity Compliance}
\label{tab:llm_compliance_comparison}

\resizebox{\linewidth}{!}{%
\begin{tabular}{p{5cm} p{5.0cm} p{3.0cm}}
\toprule

\textbf{Citation} &
\textbf{Primary Task} &
\textbf{AI / LLM} \\
\midrule

Work in progress: Leveraging LLMs for cybersecurity compliance \cite{11129535} &
Statement of Applicability (SOA) generation, audit
scope definition, and compliance checklist development &
 Sonnet 3.5, ChatGPT 4o, and Llama 3.1 \\

\addlinespace

PolicyGPT: Automated Analysis of Privacy Policies with LLMs \cite{tang2023policygpt} &
Specific to classification of privacy policies &
ChatGPT, GPT-4, and claude2 \\

\addlinespace

Enhancing privacy policy comprehension through Privacify: A user-centric approach using advanced language models \cite{WOODRING2024103997} &
To enhance the accessibility and understandability of privacy policies &
Mistral 7B Instruct \\

\addlinespace

PARAG: Proactive Answering Framework using Retrieval-Augmented Generation \cite{saha2025parag} &
Proactive Q\&A; compliance guidance retrieval &
GPT-4o-mini, Mistral-7B \\

\addlinespace

AutoPGT: LLM-Driven Automated Policy Generation for Securing Industrial Control Systems \cite{11204572} &
Automated security policy generation for ICS environments &
GPT-4o Mini, Deepseek-chat \\

\addlinespace

Position Paper: Leveraging LLMs for Cybersecurity Compliance \cite{10628537} &
Conceptual framing; outline a comprehensive roadmap for studying the utility of LLMs in cybersecurity compliance &
No implementation \\

MACAW: Automating Organizational Cyber Security Policy Compliance Against Industry Standards Using Agentic AI \cite{negi2026automating} & Policy compliance analysis using closed source models & GPT-4o Mini\\

Automated Information Security Policy Analysis:
A Conceptual Framework \cite{priescu2026automated} & automatic assessment of information security policies against the ISO/IEC 27001:2022 standard. & BERT, DistilBERT, ALBERT, RoBERTA, T5, Falcon, and BART \\

ReCon & Automatic assessment of information/cyber security policy documents against the ISO/IEC 27002:2022 standard without requiring the ISO document, using LLM prior knowledge & Open Weight CPU-Deployable LLM's \\

\bottomrule
\end{tabular}
}
\end{table}

However, using LLMs in an on-premises environment often requires high computational resources, especially GPU infrastructure. Since it is not necessary for every organization to have GPU-based resources available, low-compute solutions become increasingly important. On the other hand, models with relatively few parameters can be run on standard CPUs. The context window of such models is a crucial factor in their use, as the limited context window also limits the size of the prompt. Consequently, careful use of each token used in the prompt becomes essential.

The primary motivation of this work is to develop an automated compliance gap analysis approach that is accessible to organizations of all sizes and resource levels. By designing a solution capable of operating efficiently on standard commercial computing systems, potentially relying solely on CPU resources, we aim to minimize dependence on expensive cloud-based APIs and specialized hardware infrastructure. Such an approach enables cybersecurity teams to identify compliance deficiencies more rapidly, update policies effectively, and maintain secure IT environments while keeping operational costs low. Additionally, existing research lacks a comprehensive benchmark on the use of LLMs leveraging the prior knowledge of the LLM itself for cybersecurity policy compliance assessment, as well as standardized performance evaluations and comparative analyses across different combinations of rerankers and embedding models. By addressing these limitations, this study seeks to advance the practicality, affordability, and widespread adoption of LLM-based compliance analysis solutions.

\section{Methodology}

In this section we describe the design and key components of methodology. It analyzes an organization’s cybersecurity policies and procedures, validates them against the security standards such as ISO 27002:2022 standard, determines if relevant controls have been properly addressed by the organization’s policies, identifies any gaps in the policies, and generates easy to understand and interpretable justifications in natural language of the alignments and gaps between the policies and standard controls. We utilize RAG (Retrieval-Augmented Generation) for capturing the organization’s policies and procedures and intelligent retrieval of relevant policy information to build appropriate context for each standard cybersecurity control query. Using LLM for reasoning, improvement of decisions, and generation of justification and gap text. The overall high-level architecture of the proposed framework is illustrated in Figure \ref{fig:architecture}. The key components include:

\begin{figure}
    \centering
    \includegraphics[scale=0.55]{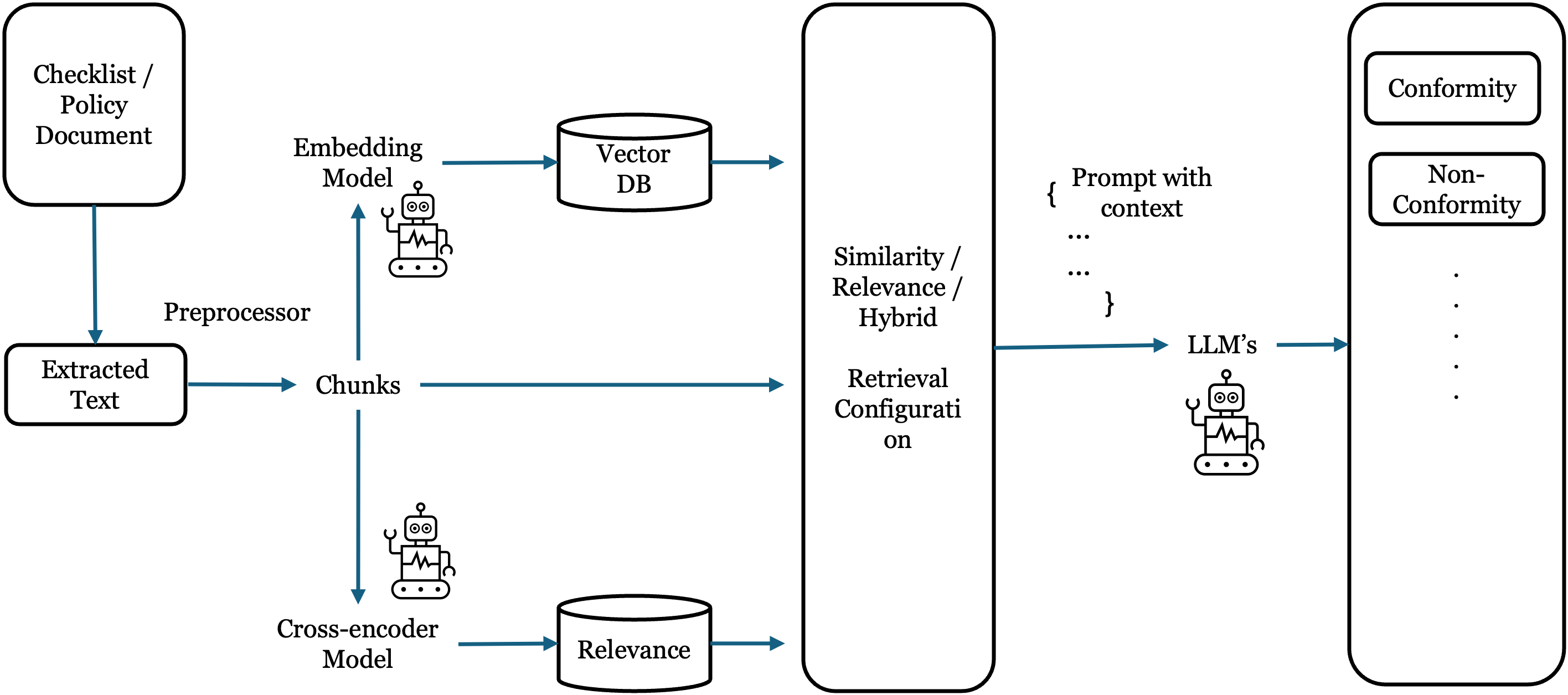}
    \caption{"ReCon" Framework Architecture and Components}
    \label{fig:architecture}
\end{figure}

\subsection{Pre-processor of cybersecurity policies}

The Pre-processor component ingests checklist data and policy documents from that can be in various file formats such as .pdf, .docx and .csv (checklist). We have used the tiktoken tokenizer \cite{openai_tiktoken} with GPT-4 encoding. We first load the document and extract each paragraph as a separate string, which gives us the raw text blocks. We then clean this text by removing empty paragraphs, normalizing whitespace, and filtering out non-ASCII characters. After cleaning, we count the tokens for each strings.

\subsection{Chunking}

For chunking, we have used the generated tokens to split the text into smaller overlapping sections for embedding. We accumulate paragraphs into a buffer until adding another would exceed a defined token limit. At that point, we store the buffer as a chunk and carry forward a small overlap into the next chunk. This overlap ensures continuity of information across chunks.

\subsection{Embedding \& Reranking}

Vector embedding: We have used an embedding model to convert each chunk generated into a numerical representation. Each chunk of text is passed through the selected embedding model (as listed in Table \ref{tab:runs}), which transforms it into a vector. These vectors are then stored in ChromaDB \cite{chroma2026}, a vector database that allows efficient storage and retrieval.


Reranking: An approach is used to identify the relevance of chunks for each compliance clause in the checklist. Each clause–chunk pair evaluated using a reranker such as the BAAI/bge-reranker-base model for relevance assessment and assigns a relevance score based on their semantic relationship.

\subsection{Intelligent retrieval \& Prompt Queue Builder}


%


\begin{figure}
    \centering
    \includegraphics[width=0.9\linewidth]{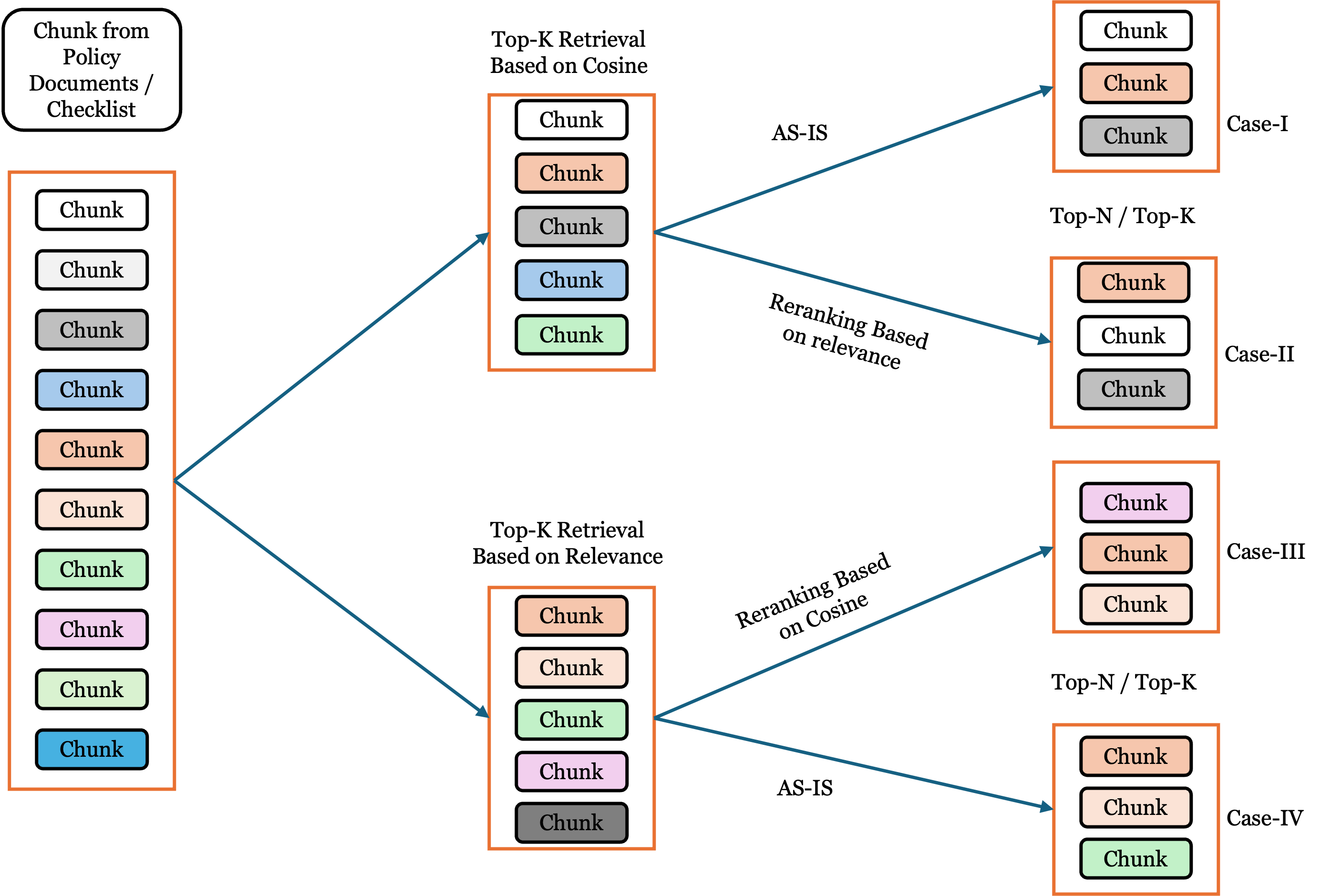}
    \caption{Four retrieval strategies}
    \label{fig:cases}
\end{figure}

The prompt has been constructed using the retrieval strategy as depicted in Figure \ref{fig:cases}. In the retrieval strategy, four cases are defined. For Case 1, we retrieve the top-K results based on similarity and then select the top-N from these. In Case 2, we rerank the top-K results using a cross-encoder and then select the top-N. In Case 3, we retrieve the top-K results based on relevance, rerank them using similarity, and then select the top-N. In Case 4, we retrieve the top-K results based on relevance. As we are using multiple embedding models and multiple cross-encoders, Table \ref{tab:runs} presents twelve unique retrieval configurations, each determining how the selected chunks are organized. These strategies allow us to experiment with different ways of providing context and to evaluate the agreement of the model’s responses.

\subsection{LLM for reasoning and compliance assessment}

For LLM reasoning and compliance assessment, we have used the Ollama tool \cite{ollama2026} for local deployment of models on the Ubuntu operating system. The deployment is designed to work in a low-compute environment, where CPU-based execution is sufficient for running selected models without requiring dedicated GPU resources. The focus is on enabling local experimentation and evaluation of reasoning and compliance behaviors in a cost-effective manner with default configurations.

\begin{table}[!t]
\centering
\caption{Retrieval configurations across embedding and reranker combinations.}
\label{tab:runs}

\small
\renewcommand{\arraystretch}{1.1}
\setlength{\tabcolsep}{4pt}

\begin{tabular}{cp{5.2cm}l}
\toprule
\textbf{Case} & \textbf{Embedding} & \textbf{Reranker} \\
\midrule

\multirow{2}{*}{1}
& BAAI/bge-base-en-v1.5 \cite{bge_base_en_v15} & -- \\ \cline{2-3}
& all-MiniLM-L6-v2 \cite{all_minilm_l6_v2} & -- \\ 
\midrule

\multirow{4}{*}{2}
& BAAI/bge-base-en-v1.5 & cross-encoder/ms-marco-MiniLM-L-12-v2 \cite{msmarco_minilm_l12_v2} \\ \cline{2-3}
& BAAI/bge-base-en-v1.5 & BAAI/bge-reranker-base \cite{bge_reranker_base} \\ \cline{2-3}
& all-MiniLM-L6-v2 & cross-encoder/ms-marco-MiniLM-L-12-v2 \\ \cline{2-3}
& all-MiniLM-L6-v2 & BAAI/bge-reranker-base \\
\midrule

\multirow{4}{*}{3}
& BAAI/bge-base-en-v1.5 & cross-encoder/ms-marco-MiniLM-L-12-v2 \\ \cline{2-3}
& BAAI/bge-base-en-v1.5 & BAAI/bge-reranker-base \\ \cline{2-3}
& all-MiniLM-L6-v2 & cross-encoder/ms-marco-MiniLM-L-12-v2 \\ \cline{2-3}
& all-MiniLM-L6-v2 & BAAI/bge-reranker-base \\
\midrule

\multirow{2}{*}{4}
& -- & cross-encoder/ms-marco-MiniLM-L-12-v2 \\ \cline{2-3}
& -- & BAAI/bge-reranker-base \\
\midrule

\multicolumn{3}{l}{\textbf{Summary of experimental scale}} \\
\midrule

\multicolumn{2}{l}{Retrieval configurations (Unique)} & 12 \\
\multicolumn{2}{l}{LLM models} & 8 \\
\multicolumn{2}{l}{Policy datasets} & 3 \\
\midrule

\multicolumn{2}{l}{Total experimental runs} & $12 \times 8 \times 3 = 288$ \\
\multicolumn{2}{l}{Controls per run} & 93 \\
\midrule

\multicolumn{2}{l}{Total evaluations (Unique)} & $288 \times 93 = 26{,}784$ \\
\bottomrule
\end{tabular}

\end{table}

\subsection{Evaluation Metrics}
The performance of each large language model (LLM) was evaluated by comparing its response with the corresponding human annotations, which served as the ground truth. Since the policy datasets exhibit substantial class imbalance, conventional classification accuracy alone is not an appropriate measure of model performance. Specifically, one dataset (Set-B) consists entirely of conforming policies (100\% conformity), while the other datasets contain approximately 59.13\% (Set-C) and 22.58\% (Set-A) conforming policies, respectively. Owing to this imbalance, the evaluation focused on measuring the agreement  between LLM predictions and human annotations rather than relying solely on overall classification performance. Accordingly, four complementary metrics were used to assess model performance: Agreement, Cohen's Kappa, Matthews Correlation Coefficient (MCC), and Balanced Accuracy.

\paragraph{Agreement Observed ($A_{o}$):}
($A_{o}$) or ``agreement" or ``accuracy" represents the proportion of evaluations in which the model response matched the human decision. In this study, agreement  is defined as the percentage of evaluations for which both the human and the model assigned the same label (i.e., \textit{conformity} or \textit{nonconformity}). It is calculated as

\[
A_o =  \mathrm{Accuracy} = \frac{TP+TN}{N}
\]

\paragraph{Cohen's Kappa ($\kappa$):}
It measures the level of agreement ($A_{o}$) between the model predictions and the human annotations after correcting for ``agreement expected by chance ($A_{e}$)". The statistic ranges from $-1$ to $1$, where a value of $1$ indicates perfect agreement ($A_{o}$), $0$ indicates agreement ($A_{o}$) equivalent to chance, and negative values indicate agreement ($A_{o}$) worse than chance.

\[
A_e =
\frac{
(TP+FP)(TP+FN) + (FN+TN)(FP+TN)
}{
N^2
}
\]

\[
\kappa =
\frac{A_o-A_e}{1-A_e}
\]

\paragraph{Matthews Correlation Coefficient ($\mathrm{MCC}$):}
It is a balanced measure of binary classification performance that incorporates true positives, true negatives, false positives, and false negatives. Unlike overall accuracy, MCC remains informative even when the class distribution is imbalanced. Its values range from $-1$ to $1$, where $1$ indicates perfect prediction, $0$ indicates performance no better than random prediction, and $-1$ indicates complete disagreement.

\[
\mathrm{MCC}
=
\frac{TP \times TN - FP \times FN}
{\sqrt{(TP+FP)(TP+FN)(TN+FP)(TN+FN)}}
\]

\paragraph{Balanced Accuracy ($\mathrm{BA}$):}
It is used to evaluate classification performance while accounting for class imbalance. It is defined as the average of sensitivity (true positive rate) and specificity (true negative rate):

\[
\mathrm{BA}
=
\frac{1}{2}
\left(
\frac{TP}{TP+FN}
+
\frac{TN}{TN+FP}
\right)
\]

\section{Experiment}

We experimented with eight open-weight, low-compute, CPU-deployable language models (as listed in Table \ref{tab:llms}). All models are deployed locally using the Ollama framework. The generation temperature is set to 0 to ensure deterministic outputs, with a maximum response time constraint of 120 seconds per query. All experiments are executed in a controlled offline environment on the system equipped with an Intel Core i9-14900 CPU, capable of a maximum clock frequency of 5.5 GHz, 31 GB of RAM, integrated Intel UHD Graphics 770, and NVMe SSD storage (Phison-based, model ABSNE191TB).

\begin{table}[!t]
\centering
\caption{Language Models compatible with Ollama}
\label{tab:llms}

\small
\renewcommand{\arraystretch}{1.1}
\setlength{\tabcolsep}{5pt}

\begin{tabular}{c p{2.2cm} c c c c}
\toprule

\textbf{S.No} & \textbf{Model} & \textbf{Params} & \textbf{Size} & \textbf{Max Context} & \textbf{Default Context} \\
\midrule

1 & Llama 3.2   & 3b  & 2.0 GB & 131072 & 4096 \\
2 & Mistral     & 7b  & 4.4 GB & 32768  & 4096 \\
3 & Granite 3.3 & 8b  & 4.9 GB & 131072 & 4096 \\
4 & Granite 4.1 & 3b  & 2.1 GB & 131072 & 4096 \\
5 & Granite 4.1 & 8b  & 5.3 GB & 131072 & 4096 \\
6 & Gemma 3     & 4b  & 3.3 GB & 131072 & 4096 \\
7 & Gemma 3n    & e2b & 5.6 GB & 32768  & 4096 \\
8 & Phi-4       & 14b & 9.1 GB & 16384  & 4096 \\
\bottomrule

\end{tabular}
\end{table}

\subsection{Datasets \& Ground Truth}
We use the most recent cybersecurity controls from ISO/IEC 27002:2022, incorporating the full set of controls defined in the standard. To analyze organizational cybersecurity policies and identify potential gaps, we collected three different sets of cybersecurity policy documents from independent CISA/ISMS LA-certified auditors. All three policies are real-world documents

We also obtained their conformity assessments directly from the three certified auditors, which is treated as ground truth. As listed in Table \ref{tab:abm_results}, The first policy set (Set A) contains 21 conformities and 72 non-conformities. The second policy set (Set B) contains all conformities. The third policy set (Set C) contains 55 conformities and 38 non-conformities.

\begin{table}
\centering
\caption{Ground Truth for ISO}
\label{tab:abm_results}

\small
\renewcommand{\arraystretch}{1.05}
\setlength{\tabcolsep}{3pt}

\begin{tabular}{l c cc cc cc}
\toprule

& & \multicolumn{2}{c}{Set-A: Majority NC}
& \multicolumn{2}{c}{Set-B: All C}
& \multicolumn{2}{c}{Set-C: Majority C} \\

\cmidrule(lr){3-4}\cmidrule(lr){5-6}\cmidrule(lr){7-8}

\textbf{Clause Category} & \textbf{Controls}
& \textbf{C} & \textbf{NC}
& \textbf{C} & \textbf{NC}
& \textbf{C} & \textbf{NC} \\

\midrule

Organizational (A.5) & 37 & 6  & 31 & 37 & 0  & 23 & 14 \\
People (A.6)         & 8  & 3  & 5  & 8  & 0  & 1  & 7  \\
Physical (A.7)       & 14 & 2  & 12 & 14 & 0  & 14 & 0  \\
Technological (A.8)  & 34 & 10 & 24 & 34 & 0  & 17 & 17 \\

\midrule
\textbf{Sub-total}   & \textbf{93} & \textbf{21} & \textbf{72}
                    & \textbf{93} & \textbf{0}  & \textbf{55} & \textbf{38} \\

\bottomrule
\end{tabular}

\vspace{2pt}
{\scriptsize Note: C = Conformity, NC = Non-Conformity}

\end{table}

\subsection{Chunking, Storing, Retrieval \& Prompts Queue Generation}
For prompt preparation, the input documents are preprocessed and segmented into chunks using a token window of 250 tokens with a 50-token overlap. These chunks are subsequently processed using embedding / cross-encoder models and indexed for retrieval.

For intelligent information retrieval, both embedding-based similarity search and cross-encoder re-ranking strategies are employed. A total of 12 retrieval configurations are defined, combining different embedding and cross-encoder models as summarized in Table \ref{tab:runs}.

Along with eight language models, three policy datasets, four embedding–re-ranker combinations, and four retrieval strategies, this resulted in 384 possible combinations. However, there is redundancy; for example, the Top-N chunks retrieved using Embedding-1 without Cross Encoder-1 and Embedding-1 without Cross Encoder-2 are identical. As a result, the total number of unique experimental configurations is reduced to 288. Each configuration is evaluated using 93 controlled prompts, leading to a total of 26,784 prompt-level evaluations. These redundant runs have been used to check the consistency of the responses of the LLMs at temperature 0. For this experiment, the retrieval parameters were set to Top-K = 15 and Top-N = 5.

\subsubsection{LLM Selection}
Eight different models were selected and deployed using Ollama \cite{ollama2026}, a local inference framework for running open-weight LLMs on consumer hardware, as listed in Table \ref{tab:llms}. The selection was motivated by their accessibility \& cost-effectiveness.

The choice of models was informed by the study conducted by \cite{negi2026small}, which benchmarked 16 low-compute LLMs based on their ability to recall guidance associated with the 93 controls of ISO 27002:2022. From these 16 models, we selected the six highest-performing models in terms of coverage: phi4:14b (83\%), gemma3:4b (80.66\%), granite4.1:3b (80.26\%), gemma3n:e2b (79.72\%), granite4.1:8b (78.25\%) and mistral:7b (78.10\%). In addition, we included two representative models that achieved coverage above 70\%, namely , granite3.3:3b, and llama3.2:3b.

\section{Results}
\label{sec:results}

This section presents the benchmarking results for eight LLMs evaluated as automated ISMS Lead Auditor across 93 ISO/IEC 27002:2022 controls. Each model received a structured prompt specifying audit role, evaluation criteria, output format, and policy evidence, and produced a JSON verdict of CONFORMITY or NON-CONFORMITY and justification.

\subsection{Overall Model Performance}

Table \ref{tab:model_performance} presents the performance comparison of the selected LLMs. Among all evaluated models, Granite4.1:8B achieved the best overall performance with a balanced accuracy of 75.0\%, MCC of 0.548, and Cohen's Kappa of 0.528. All variants of Granite demonstrated competitive performance followed by llama3.2. Remaining all models has a balanced accuracy below than 60\%. 

\begin{table}[ht]
\centering
\caption{Performance comparison of different language models using Balanced Accuracy, MCC, and Cohen's Kappa.}
\label{tab:model_performance}
\begin{tabular}{p{2cm} ccc}
\hline
\textbf{Model} & \textbf{Balanced Accuracy (\%)} & \textbf{MCC} & \textbf{Cohen's Kappa} \\
\hline
Gemma3          & 54.3 & 0.145 & 0.069 \\
Gemma3n         & 52.4 & 0.094 & 0.038 \\
Granite3.3      & 68.0 & 0.462 & 0.398 \\
Granite4.1:3B   & 73.4 & 0.460 & 0.458 \\
Granite4.1:8B   & 75.0 & 0.548 & 0.528 \\
Llama3.2        & 64.4 & 0.328 & 0.308 \\
Mistral         & 54.6 & 0.210 & 0.109 \\
Phi4            & 59.9 & 0.291 & 0.163 \\
\hline
\end{tabular}
\end{table}

\subsection{Mean Agreement analysis with Ground truth}

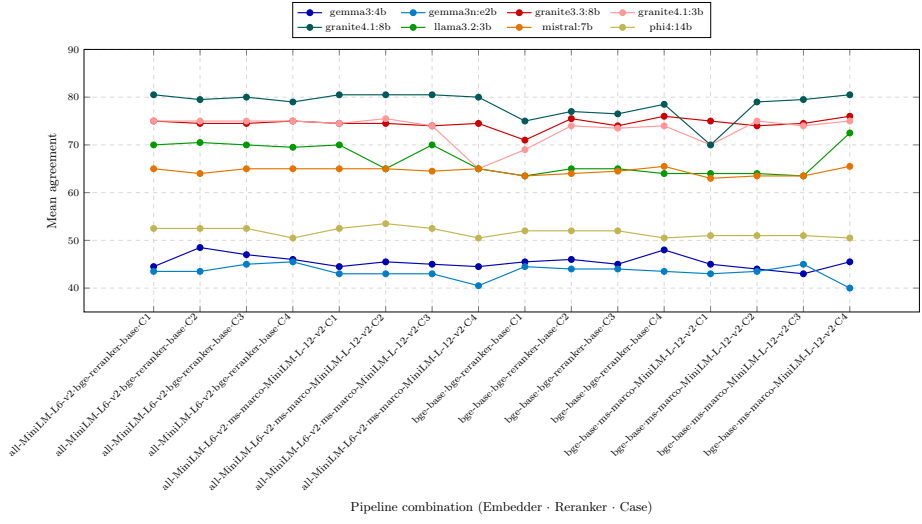
\begin{figure}[h]
\centering
\resizebox{\textwidth}{!}{%
\begin{tikzpicture}
\begin{axis}[
    width=22cm,
    height=8cm,
    xlabel={Pipeline combination (Embedder $\cdot$ Reranker $\cdot$ Case)},
    ylabel={Mean agreement},
    ymin=35, ymax=90,
    xtick=data,
    xticklabels={
        {all-MiniLM-L6-v2$\cdot$bge-reranker-base$\cdot$C1},
        {all-MiniLM-L6-v2$\cdot$bge-reranker-base$\cdot$C2},
        {all-MiniLM-L6-v2$\cdot$bge-reranker-base$\cdot$C3},
        {all-MiniLM-L6-v2$\cdot$bge-reranker-base$\cdot$C4},
        {all-MiniLM-L6-v2$\cdot$ms-marco-MiniLM-L-12-v2$\cdot$C1},
        {all-MiniLM-L6-v2$\cdot$ms-marco-MiniLM-L-12-v2$\cdot$C2},
        {all-MiniLM-L6-v2$\cdot$ms-marco-MiniLM-L-12-v2$\cdot$C3},
        {all-MiniLM-L6-v2$\cdot$ms-marco-MiniLM-L-12-v2$\cdot$C4},
        {bge-base$\cdot$bge-reranker-base$\cdot$C1},
        {bge-base$\cdot$bge-reranker-base$\cdot$C2},
        {bge-base$\cdot$bge-reranker-base$\cdot$C3},
        {bge-base$\cdot$bge-reranker-base$\cdot$C4},
        {bge-base$\cdot$ms-marco-MiniLM-L-12-v2$\cdot$C1},
        {bge-base$\cdot$ms-marco-MiniLM-L-12-v2$\cdot$C2},
        {bge-base$\cdot$ms-marco-MiniLM-L-12-v2$\cdot$C3},
        {bge-base$\cdot$ms-marco-MiniLM-L-12-v2$\cdot$C4}
    },
    x tick label style={rotate=45, anchor=east, font=\scriptsize},
    legend style={
        at={(0.5,1.18)},
        anchor=north,
        legend columns=4,
        font=\scriptsize,
        /tikz/every even column/.append style={column sep=0.5em}
    },
    grid=major,
    grid style={dashed, gray!30},
    tick label style={font=\scriptsize},
    label style={font=\small},
    mark size=2pt,
    cycle list name=color list,
]

\addplot[color=blue!70!black, mark=*, thick] coordinates {
    (1,44.5)(2,48.5)(3,47)(4,46)(5,44.5)(6,45.5)(7,45)(8,44.5)
    (9,45.5)(10,46)(11,45)(12,48)(13,45)(14,44)(15,43)(16,45.5)
};
\addlegendentry{gemma3:4b}

\addplot[color=cyan!60!blue, mark=*, thick] coordinates {
    (1,43.5)(2,43.5)(3,45)(4,45.5)(5,43)(6,43)(7,43)(8,40.5)
    (9,44.5)(10,44)(11,44)(12,43.5)(13,43)(14,43.5)(15,45)(16,40)
};
\addlegendentry{gemma3n:e2b}

\addplot[color=red!80!black, mark=*, thick] coordinates {
    (1,75)(2,74.5)(3,74.5)(4,75)(5,74.5)(6,74.5)(7,74)(8,74.5)
    (9,71)(10,75.5)(11,74)(12,76)(13,75)(14,74)(15,74.5)(16,76)
};
\addlegendentry{granite3.3:8b}

\addplot[color=red!40!white, mark=*, thick] coordinates {
    (1,75)(2,75)(3,75)(4,75)(5,74.5)(6,75.5)(7,74)(8,65)
    (9,69)(10,74)(11,73.5)(12,74)(13,70)(14,75)(15,74)(16,75)
};
\addlegendentry{granite4.1:3b}

\addplot[color=teal!70!black, mark=*, thick] coordinates {
    (1,80.5)(2,79.5)(3,80)(4,79)(5,80.5)(6,80.5)(7,80.5)(8,80)
    (9,75)(10,77)(11,76.5)(12,78.5)(13,70)(14,79)(15,79.5)(16,80.5)
};
\addlegendentry{granite4.1:8b}

\addplot[color=green!60!black, mark=*, thick] coordinates {
    (1,70)(2,70.5)(3,70)(4,69.5)(5,70)(6,65)(7,70)(8,65)
    (9,63.5)(10,65)(11,65)(12,64)(13,64)(14,64)(15,63.5)(16,72.5)
};
\addlegendentry{llama3.2:3b}

\addplot[color=orange!90!black, mark=*, thick] coordinates {
    (1,65)(2,64)(3,65)(4,65)(5,65)(6,65)(7,64.5)(8,65)
    (9,63.5)(10,64)(11,64.5)(12,65.5)(13,63)(14,63.5)(15,63.5)(16,65.5)
};
\addlegendentry{mistral:7b}

\addplot[color=yellow!70!black, mark=*, thick] coordinates {
    (1,52.5)(2,52.5)(3,52.5)(4,50.5)(5,52.5)(6,53.5)(7,52.5)(8,50.5)
    (9,52)(10,52)(11,52)(12,50.5)(13,51)(14,51)(15,51)(16,50.5)
};
\addlegendentry{phi4:14b}

\end{axis}
\end{tikzpicture}
}
\caption{Mean of agreement of all evaluated LLMs across 16 pipeline combinations (embedder $\times$ reranker $\times$ case).}
\label{fig:all-models-pipeline}
\end{figure}

We calculated the mean agreement for each retrieval strategy and observed that all variants of Granite consistently achieved higher agreement scores than the other models. As depicted in Figure \ref{fig:all-models-pipeline}, Granite 4.1:8B performed the best, achieving an agreement of 80.65\%, followed by Granite 4.1:3B (76.34\%) and Granite 3.3:8B (75.98\%). The next best-performing model was Llama 3.2:3B, with an agreement of 71.32\%, followed by Mistral:7B, which achieved 65.59\% agreement.


\subsection{Contribution of Retrieval Strategy (Case)}
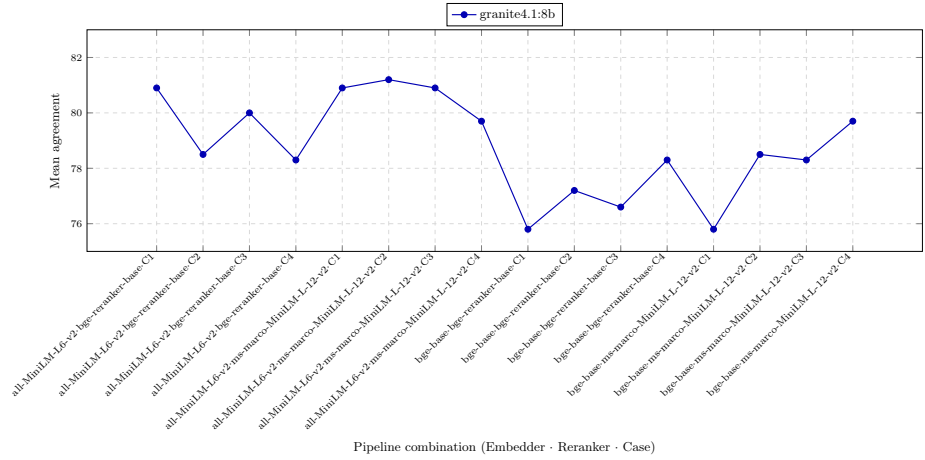
\begin{figure}[h]
\centering
\resizebox{\textwidth}{!}{%
\begin{tikzpicture}
\begin{axis}[
    width=22cm,
    height=7cm,
    xlabel={Pipeline combination (Embedder $\cdot$ Reranker $\cdot$ Case)},
    ylabel={Mean agreement},
    ymin=75, ymax=83,
    xtick=data,
    xticklabels={
        {all-MiniLM-L6-v2$\cdot$bge-reranker-base$\cdot$C1},
        {all-MiniLM-L6-v2$\cdot$bge-reranker-base$\cdot$C2},
        {all-MiniLM-L6-v2$\cdot$bge-reranker-base$\cdot$C3},
        {all-MiniLM-L6-v2$\cdot$bge-reranker-base$\cdot$C4},
        {all-MiniLM-L6-v2$\cdot$ms-marco-MiniLM-L-12-v2$\cdot$C1},
        {all-MiniLM-L6-v2$\cdot$ms-marco-MiniLM-L-12-v2$\cdot$C2},
        {all-MiniLM-L6-v2$\cdot$ms-marco-MiniLM-L-12-v2$\cdot$C3},
        {all-MiniLM-L6-v2$\cdot$ms-marco-MiniLM-L-12-v2$\cdot$C4},
        {bge-base$\cdot$bge-reranker-base$\cdot$C1},
        {bge-base$\cdot$bge-reranker-base$\cdot$C2},
        {bge-base$\cdot$bge-reranker-base$\cdot$C3},
        {bge-base$\cdot$bge-reranker-base$\cdot$C4},
        {bge-base$\cdot$ms-marco-MiniLM-L-12-v2$\cdot$C1},
        {bge-base$\cdot$ms-marco-MiniLM-L-12-v2$\cdot$C2},
        {bge-base$\cdot$ms-marco-MiniLM-L-12-v2$\cdot$C3},
        {bge-base$\cdot$ms-marco-MiniLM-L-12-v2$\cdot$C4}
    },
    x tick label style={rotate=45, anchor=east, font=\scriptsize},
    legend style={
        at={(0.5,1.12)},
        anchor=north,
        legend columns=1,
        font=\small,
    },
    grid=major,
    grid style={dashed, gray!30},
    tick label style={font=\scriptsize},
    label style={font=\small},
    mark size=2pt,
]

\addplot[color=blue!70!black, mark=*, thick] coordinates {
    (1,80.9)(2,78.5)(3,80.0)(4,78.3)(5,80.9)(6,81.2)(7,80.9)(8,79.7)
    (9,75.8)(10,77.2)(11,76.6)(12,78.3)(13,75.8)(14,78.5)(15,78.3)(16,79.7)
};
\addlegendentry{granite4.1:8b}

\end{axis}
\end{tikzpicture}
}
\caption{granite4.1:8b mean agreement across all 16 pipeline combinations.}
\label{fig:granite41-8b-zoomed}
\end{figure}

Four retrieval strategies were evaluated: CE (cross-encoder only), CE→Cosine (cross-encoder retrieval followed by cosine re-scoring), Cosine→CE (cosine retrieval followed by cross-encoder re-scoring), and Cosine (cosine similarity only). As depicted in Figure~\ref{fig:granite41-8b-zoomed}, the highest mean agreement across all 16 pipeline combinations for Granite 4.1:8B (80.64\%) was achieved using the all-MiniLM-L6-v2 with ms-marco-MiniLM-L-12-v2 (Case 2) pipeline. In contrast, performance was notably lower (75.62\%) when using the bge-base embedder with the bge-reranker-base re-ranker (Case 1) and the \texttt{ms-marco} re-ranker (Case 1). This analysis helps identify the embedding model and re-ranker combination that achieves the highest agreement.

\subsection{Hardware Resource Benchmarks}

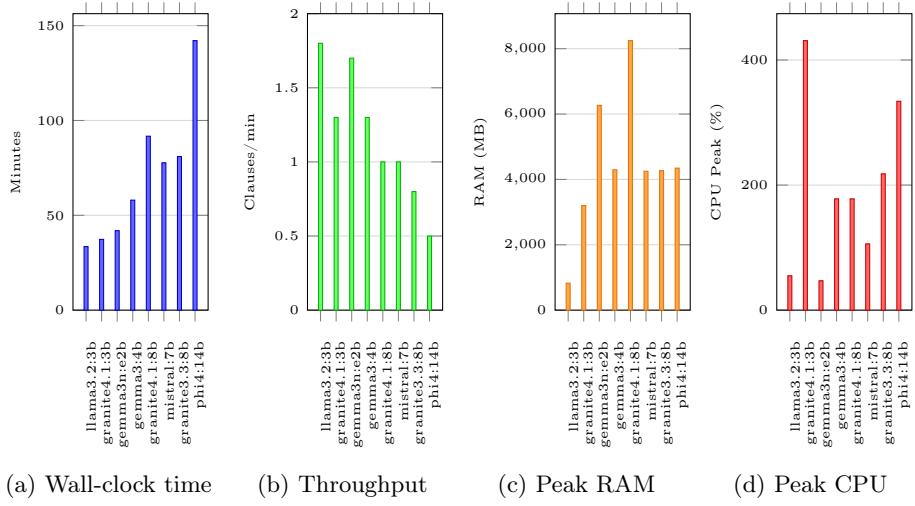
\begin{figure}[h]
\centering
\begin{subfigure}[t]{0.24\textwidth}
\centering
\begin{tikzpicture}
\begin{axis}[
    ybar,
    width=1.15\linewidth,
    height=5.5cm,
    ymin=0,
    ylabel={Minutes},
    bar width=1.5pt,
    enlarge x limits=0.12,
    xtick=data,
    symbolic x coords={
        llama3.2:3b,
        granite4.1:3b,
        gemma3n:e2b,
        gemma3:4b,
        granite4.1:8b,
        mistral:7b,
        granite3.3:8b,
        phi4:14b
    },
    x tick label style={
        rotate=90,
        anchor=east,
        font=\fontsize{5}{6}\selectfont,
        xshift=-2pt,
        yshift=-2pt
    },
    y tick label style={font=\fontsize{5}{6}\selectfont},
    ylabel style={font=\fontsize{5}{6}\selectfont},
    ymajorgrids,
    grid style={gray!30},
    xtick align=outside,
    xticklabel shift=2pt,
]
\addplot[fill=blue!60, draw=blue!80!black] coordinates {
    (llama3.2:3b,33.5)
    (granite4.1:3b,37.3)
    (gemma3n:e2b,41.9)
    (gemma3:4b,58.0)
    (granite4.1:8b,91.7)
    (mistral:7b,77.7)
    (granite3.3:8b,81.0)
    (phi4:14b,142.1)
};
\end{axis}
\end{tikzpicture}
\caption{Wall-clock time}
\end{subfigure}
\hfill
\begin{subfigure}[t]{0.24\textwidth}
\centering
\begin{tikzpicture}
\begin{axis}[
    ybar,
    width=1.15\linewidth,
    height=5.5cm,
    ymin=0,
    ymax=2,
    ylabel={Clauses/min},
    bar width=1.5pt,
    enlarge x limits=0.12,
    xtick=data,
    symbolic x coords={
        llama3.2:3b,
        granite4.1:3b,
        gemma3n:e2b,
        gemma3:4b,
        granite4.1:8b,
        mistral:7b,
        granite3.3:8b,
        phi4:14b
    },
    x tick label style={
        rotate=90,
        anchor=east,
        font=\fontsize{5}{6}\selectfont,
        xshift=-2pt,
        yshift=-2pt
    },
    y tick label style={font=\fontsize{5}{6}\selectfont},
    ylabel style={font=\fontsize{5}{6}\selectfont},
    ymajorgrids,
    grid style={gray!30},
    xtick align=outside,
    xticklabel shift=2pt,
]
\addplot[fill=green!60, draw=green!70!black] coordinates {
    (llama3.2:3b,1.8)
    (granite4.1:3b,1.3)
    (gemma3n:e2b,1.7)
    (gemma3:4b,1.3)
    (granite4.1:8b,1.0)
    (mistral:7b,1.0)
    (granite3.3:8b,0.8)
    (phi4:14b,0.5)
};
\end{axis}
\end{tikzpicture}
\caption{Throughput}
\end{subfigure}
\hfill
\begin{subfigure}[t]{0.24\textwidth}
\centering
\begin{tikzpicture}
\begin{axis}[
    ybar,
    width=1.15\linewidth,
    height=5.5cm,
    ymin=0,
    ylabel={RAM (MB)},
    bar width=1.5pt,
    enlarge x limits=0.12,
    xtick=data,
    symbolic x coords={
        llama3.2:3b,
        granite4.1:3b,
        gemma3n:e2b,
        gemma3:4b,
        granite4.1:8b,
        mistral:7b,
        granite3.3:8b,
        phi4:14b
    },
    x tick label style={
        rotate=90,
        anchor=east,
        font=\fontsize{5}{6}\selectfont,
        xshift=-2pt,
        yshift=-2pt
    },
    y tick label style={font=\fontsize{5}{6}\selectfont},
    ylabel style={font=\fontsize{5}{6}\selectfont},
    ymajorgrids,
    grid style={gray!30},
    xtick align=outside,
    xticklabel shift=2pt,
]
\addplot[fill=orange!70, draw=orange!90!black] coordinates {
    (llama3.2:3b,825)
    (granite4.1:3b,3200)
    (gemma3n:e2b,6266)
    (gemma3:4b,4295)
    (granite4.1:8b,8247)
    (mistral:7b,4249)
    (granite3.3:8b,4268)
    (phi4:14b,4345)
};
\end{axis}
\end{tikzpicture}
\caption{Peak RAM}
\end{subfigure}
\hfill
\begin{subfigure}[t]{0.24\textwidth}
\centering
\begin{tikzpicture}
\begin{axis}[
    ybar,
    width=1.15\linewidth,
    height=5.5cm,
    ymin=0,
    ylabel={CPU Peak (\%)},
    bar width=1.5pt,
    enlarge x limits=0.12,
    xtick=data,
    symbolic x coords={
        llama3.2:3b,
        granite4.1:3b,
        gemma3n:e2b,
        gemma3:4b,
        granite4.1:8b,
        mistral:7b,
        granite3.3:8b,
        phi4:14b
    },
    x tick label style={
        rotate=90,
        anchor=east,
        font=\fontsize{5}{6}\selectfont,
        xshift=-2pt,
        yshift=-2pt
    },
    y tick label style={font=\fontsize{5}{6}\selectfont},
    ylabel style={font=\fontsize{5}{6}\selectfont},
    ymajorgrids,
    grid style={gray!30},
    xtick align=outside,
    xticklabel shift=2pt,
]
\addplot[fill=red!60, draw=red!80!black] coordinates {
    (llama3.2:3b,55)
    (granite4.1:3b,431)
    (gemma3n:e2b,47)
    (gemma3:4b,178)
    (granite4.1:8b,178)
    (mistral:7b,106)
    (granite3.3:8b,218)
    (phi4:14b,334)
};
\end{axis}
\end{tikzpicture}
\caption{Peak CPU}
\end{subfigure}

\caption{Hardware resource benchmark across evaluated LLMs: (a) wall-clock execution time, (b) throughput, (c) peak RAM, and (d) peak CPU utilization.}
\label{fig:hardwareBenchmark}
\end{figure}

Figure \ref{fig:hardwareBenchmark} presents hardware resource utilisation per model, measured on a CPU-only x86\_64 machine. We benchmarked each model in terms of throughput (classifications per minute), wall-clock execution time, CPU utilization, and peak RAM usage.

We observed that Llama3.2:3b demonstrated the highest throughput with the lowest memory requirement. However, Granite 4.1:8b required the highest memory, Granite 4.1:3b consumed the highest computational resources (i.e. $>$ 400\% which means all four cores saturated at 100\%), and Phi-4:14b exhibited the highest response time. These observations indicate that different models exhibit different performance characteristics across various evaluation metrics.

\subsection{Error Analysis}
Since the recommended prompt structure varies across models, we used a generic prompt, which is demonstrated in the sample available in the Appendix. For the example, we chose two clauses (i.e. A.5.11 \& A.8.11). In the first example, our ground truth is Conformity, while in the second example, the ground truth is Non-Conformity. According to our ground truth, the result for A.5.11 is Conformity, while the result for A.8.11 is Non-Conformity.

In this example, we have shown how different models respond if the prompt is generated based on the Top-k = 5 retrieved chunks for the relevant clause.

The analysis found that Mistral responded Conformity to both clauses, while the second example was actually Non-Conformity. Llama3.2:3b, on the other hand, when given a relatively loose prompt, repeated the question. By making the prompt more strict, Llama3.2:3b provided properly structured answers, but did not include the evidence that was explicitly requested. Additionally, a significant effect of the strict prompt was that it became easier to answer for relatively weak models because the prompt contained the instruction "when uncertain, then Non-Conformity".

Another important observation was that the prompt included the text of the ISO requirement. Many models mistook this requirement text for part of the policy and presented it as if it were actually in the policy. For example, Gemma3n:e2b responded that the policy mentioned data masking, even though the actual policy did not mention it. This is not only a limitation of the model, but also a problem with the prompt design, as the relatively weak models are unable to clearly distinguish between the requirement text and policy evidence due to both being in the same context window.

Analysis of the confidence scores also revealed that the model gave a confidence score of 30 for each non-conformity answer, while giving a confidence score of 90 or 95 for each conformity answer. Specifically, Gemma3:4B gave a confidence score of 65 in both instances—once when her answer was correct and once when it was wrong. This makes it clear that the confidence score was not actually a measure of the model's certainty, but merely a reflection of its final verdict.

Granite4.1:8b and Phi4:14b, which had good coverage in previous research\cite{negi2026small}, provided relatively clean and easily checkable answers.

\section{Conclusion}

In conclusion, This work demonstrates a Resource-Constrained Benchmark for Evaluating Large Language Models Under Varying Ingestion and Retrieval Configurations. First, it demonstrates a structured approach for integrating embedding models, cross-encoders, and retrieval strategies to perform automated compliance gap analysis. By systematically evaluating 384 combinations, of which 288 represent unique model- and retrieval strategy-specific prompt configurations, with each unique configuration evaluated against 93 controls, the study produced a total of 26,784 unique prompt evaluations. An additional 98 runs, each consisting of 93 control evaluations, were repeated to validate the consistency of LLM responses when using a temperature of 0. Second, the research establishes a benchmark methodology for eight well known open weight LLM models for the compliance check automation against human expert consensus. Easing the selection of the LLM's for AI in cybersecurity. Third, the research establishes the benchmark for the selection between embedding or cross encoding or hybrid approaches and retrieval strategies. which will ease the selection of approach specific to the model. Fourth, the study highlights the practical implications of running LLM-based compliance analysis on commercially available CPU resources, providing insights into the trade-offs between computational requirements and agreement  performance. For example, organizations with limited processing capacity or memory resources may choose Llama 3.2 because of its lower computational demands. However, this choice comes with a trade-off in agreement  performance compared with larger models such as Granite4.1:8B.

The findings of this study have significant practical implications for multiple stakeholders in cybersecurity compliance. For auditors, The proposed tool enhances the scalability of the auditing process by enabling a single auditor to efficiently analyze multiple policy documents, rather than being limited to one policy set at a time. This automation reduces cognitive fatigue associated with repetitive manual review. For auditees (organizations being assessed), the system enables rapid identification of gaps in their cybersecurity policies against standards such as ISO 27002:2022, allowing timely remediation and proactive management of compliance risks. The tool can be installed locally at the auditee’s site, allowing auditors to run assessments in front of the organization without any policy data leaving their environment. Auditees can also use this approach for audit monitoring and for comparing the auditor’s findings. For regulatory bodies, this tool enhances scalability and reduces cognitive fatigue. This facilitates better oversight, and evaluation of industry-wide adherence to cybersecurity standards. Additionally, the tool is designed to run on standard commercial computing systems without relying on costly cloud-based APIs or specialized HPC/GPU infrastructure, making advanced compliance gap analysis accessible to organizations of all sizes. Overall, the study demonstrates a cost-effective, and practical approach to improving cybersecurity governance and compliance while preserving privacy, as no data leaves the organizational premises due to the use of locally deployed open-source LLMs.

Although the framework evaluated 384 pipeline configurations, exhaustive exploration of additional parameters was kept for future work. First, all experiments were conducted using a fixed temperature of 0 to ensure deterministic and reproducible responses; the effects of varying temperature on compliance analysis accuracy and consistency needs to be investigated. Second, the retrieval pipeline employed a fixed top-(k) value, and the impact of different retrieval depths on agreement  and computational performance needs to be explored. Third, the maximum context length and token limits were held constant, and the influence of larger context windows on retrieval quality and reasoning performance need to be evaluated. Fourth, the proposed framework relies on zero-shot prompting, and the potential benefits of few-shot or chain-of-thought prompting needs to be examined. Fifth, conformity decisions were based on a fixed agreement  threshold; future work should investigate adaptive or statistically derived threshold selection methods to improve the robustness of compliance classification. Future work will focus on enhancing the system through fine-tuning, few-shot learning, and model distillation techniques to further improve assessment precision and contextual understanding.

\subsection*{Conflict of Interest}
The authors declare that they have no known competing financial interests or personal relationships that could have appeared to influence the work reported in this paper.

\subsection*{Data Availability}
The masked dataset will be made available upon reasonable request from the corresponding author.

\subsection*{Funding}
This research received no specific grant from any funding agency, commercial, or not-for-profit sectors.

\subsection*{AI-assisted copy editing}
The authors declare that AI-assisted copy editing was utilized in the preparation of this manuscript to enhance clarity and readability. AI tools only be applied with human oversight and control. However, all intellectual content, expertise, and conclusions remain solely the responsibility of the authors.

\bibliographystyle{unsrt}
\bibliography{ref}

\clearpage

\appendix

\end{document}